\documentclass[12pt,preprint]{aastex}

\usepackage{color}
\usepackage{epsfig}
\usepackage{pslatex}
\usepackage{graphicx}

\newcommand{\GALEX}{{\it GALEX}}

\newcommand{\Htwoo}{\mbox{H$_{2}$O}}

\newcommand{\cstwo}{\mbox{CS$_{2}$}}

\newcommand{\kms}{km~s$^{-1}$}

\newcommand{\IUE}{{\em IUE}}

\newcommand{\mols}{molecules~s$^{-1}$}

\shorttitle{GALEX Observations of Comet 9P/Tempel~1}
\shortauthors{Feldman et al.}

\begin{document}

\title{GALEX Observations of CS and OH Emission in Comet 9P/Tempel 1
During Deep Impact\altaffilmark{1}}

\author{Paul D. Feldman\altaffilmark{2}, Stephan R.
McCandliss\altaffilmark{2}, Jeffrey~P.~Morgenthaler\altaffilmark{3},\\
Carey M. Lisse\altaffilmark{4}, 
Harold A. Weaver\altaffilmark{4}, and Michael F. A'Hearn\altaffilmark{5}}


\altaffiltext{1}{Based on observations made with the NASA {\it Galaxy
Evolution Explorer}.  \GALEX\ is operated for NASA by the California
Institute of Technology under NASA contract NAS5-98034}
\altaffiltext{2}{Department of Physics and Astronomy, The Johns Hopkins University, Baltimore, MD 21218-2695}
\email{pdf@pha.jhu.edu}
\altaffiltext{3}{Planetary Science Institute, 1700 E. Fort Lowell, Suite 106,
Tucson, AZ 85719-2395}
\altaffiltext{4}{Johns Hopkins University Applied Physics Laboratory, 
Space Department, 11100 Johns Hopkins Road, Laurel, MD 20723-6099}
\altaffiltext{5}{Astronomy Department, University of Maryland, College Park, MD
20742-2421}


\pagestyle{myheadings}

\begin{abstract}
\GALEX\ observations of comet 9P/Tempel 1 using the near ultraviolet
(NUV) objective grism were made before, during and after the Deep
Impact event that occurred on 2005 July 4 at 05:52:03 UT when a 370 kg
NASA spacecraft was maneuvered into the path of the comet. The NUV
channel provides usable spectral information in a bandpass covering
2000 -- 3400 \AA\ with a point source spectral resolving power of $R
\approx 100$. The primary spectral features in this range include
solar continuum scattered from cometary dust and emissions from OH and
CS molecular bands centered near 3085 and 2575 \AA, respectively.  In
particular, we report the only cometary CS emission detected during
this event. The observations allow the evolution of these spectral
features to be tracked over the period of the encounter.  In general,
the NUV emissions observed from Tempel~1 are much fainter than those
that have been observed by \GALEX\ from other comets.  However, it is
possible to derive production rates for the parent molecules of the
species detected by \GALEX\ in Tempel 1 and to determine the number of
these molecules liberated by the impact.  The derived quiescent
production rates are $Q$(\Htwoo) $= 6.4 \times 10^{27}$~\mols\ and
$Q$(\cstwo) $= 6.7 \times 10^{24}$~\mols, while the impact produced an
additional $1.6 \times 10^{32}$~\Htwoo\ molecules and $1.3 \times
10^{29}$~\cstwo\ molecules, a similar ratio as in quiescent
outgassing.

\end{abstract}

\keywords{comets: individual (9P/Tempel~1) --- ultraviolet: solar system}

\newpage

\section{Introduction}

On 2005 July 4 at 05:52:03 UT (as seen from Earth) a 370 kg NASA
spacecraft was maneuvered into the path of comet 9P/Tempel 1
\citep{A'Hearn:2005}.  The results of the impact were recorded by a
host of ground and space based observatories with wavelength coverage
spanning the sub-mm to the ultraviolet \citep{Meech:2005}.  Here we
report on the results of low resolution slitless spectroscopy acquired
with the {\it Galaxy Evolution Explorer} (\GALEX) in the near ultraviolet NUV
channel 2 days before, immediately following, and 1 day after the
impact event.  The primary spectral features in this range include a
solar continuum scattered from cometary dust and emissions from OH and
CS molecular bands centered near 3085 and 2575~\AA, respectively.

\section{Observations}

\GALEX\ {\it (Galaxy Evolution Explorer)} is a NASA Small Explorer
whose primary mission is to map the history of star formation using two
modes: two-band photometry (FUV, 1350--1750 \AA; NUV, 1750--3100 \AA)
and integrated field grism spectroscopy with 10--20 \AA\ spectral
resolution.  Because of its large field-of-view, 1\fdg2, \GALEX\ is
also well suited to cometary coma studies as demonstrated by the 2005
March observations of C/2004 Q2 Machholz \citep{Morgenthaler:2006}.  It
has a limiting spatial resolution in the NUV channel of $\approx$
5\arcsec\ sampled with 1\farcs5 pixels.  For {\it Deep Impact} only NUV
data was obtained as the FUV side was not operating at time of the
event.  In all there were 7 contiguous pre-impact orbits devoted to grism
observations, with $\approx 900$~s visibility per orbit.  Two orbits
of  direct imaging were taken before and after the grism orbits.  For
the impact event there were 6 orbits of grism observations beginning a
few minutes after impact, with direct imaging before (1 orbit) and
after (2 orbits).  One day post-impact there were 4 orbits of grism
observations with one direct imaging orbit before the grism
observations.  A log of the spectroscopic observations is given in
Table~\ref{obslog}.  Here we concentrate on the results derived from
the low resolution slitless spectroscopy in the NUV channel.

\placetable{obslog}

At the time of impact comet 9P/Tempel 1 was at a distance of $r$ = 1.51
AU from the Sun and $\Delta$ = 0.89 AU from the Earth.  A 1\farcs5
pixel subtended 975 km at the comet.  Emission from OH, CS, and solar
scattered light were detected prior to impact.  Just after impact the
brightness and spatial extent of these emissions increased with time.
The following day saw the return of these emissions to near the
pre-impact levels.

\subsection{Data Extraction}

The \GALEX\ project supplied a time-tagged photon list of the
observations with photon coordinates referenced to RA and DEC (J2000).
The project also supplied the output of Emmanuel Bertin's SExtractor program
run on the direct-mode (imaging) observations centered on the same
fields.  SExtractor provides accurate astrometry for sources found in the
direct images.  We converted these positions into rectangular strips
using the GALEX dispersion relation for box length and the magnitude
estimates from SExtractor for box widths in order to mask sources in the
grism images.  All of our astrometric corrections were done on a spherical
projection with the IDLASTRO routines.  Co-alignment between the strips
and the grism images of the stars across the entire image confirmed the
accuracy of these calculations.  We used the JPL HORIZONS ephemeris
generator with \GALEX\ as the observatory ("@galex") to calculate the
astrometrically correct ephemeris for comet Tempel 1.  The ephemeris was
used to reconstruct the comet image from the photon list in the
reference frame of the comet.  Counts and exposure times were accumulated
into separate images and the source mask and flat field (supplied by
the \GALEX\ project) were applied
every time the stars moved by one pixel relative to the comet.  As a
result, emission from well-characterized sources was effectively erased
and very little exposure time was lost.  The final rate images,
constructed by dividing the total count images by the total exposure
time images for the corresponding observations, were used in the
analysis of the temporal evolution of the coma following the impact.

Because of the low total counts in each individual grism image,
individual exposures were summed to produce a composite image from each
day of the program.  In this case the source mask was not applied and the
stellar spectra appear as multiple images separated by the comet's
motion between \GALEX\ orbits.  An example is shown in
Figure~\ref{image} for the sum of the six grism images obtained
following the impact event.  Note that the stellar images (each
corresponding to a different orbit) are broadened by their apparent
relative motion nearly normal to the dispersion direction during the
exposure.

\placefigure{image}

\section{Analysis}

Analysis of the spectrum is begun with the composite images (such as in
Figure~\ref{image}) corresponding to the \GALEX\ orbits from the three
dates of observation.  The spectra are extracted over 13 rows of the
rotated image array, corresponding to an effective slit height of
$19.\!''5$ to ensure that only pixels with statistically significant
counts are included in the aperture photometry.  The detector
background is evaluated over the same rows in the spectral range 1700
to 2000~\AA\ where the NUV throughput for a solar-like spectrum is
effectively zero.  This background is subject to variation due to the
presence of faint stars or galaxies in the field.  The three composite
spectra, after subtraction of background, are shown in the top panels
of Figure~\ref{spec}.  The extraction of the discrete emissions from CS
and OH is complicated by the blended continuous solar spectrum produced
by the scattering of sunlight by cometary dust grains in the coma.  An
additional complication arises from the different spatial distributions
of the dust and gas components that result from the quiescent outflow
and the subsequent time evolution of these distributions after the
impact, as the spatial variation is superimposed on the actual spectrum
in the dispersion direction.

\placefigure{spec}

Between 2000 and 3400~\AA\ the solar spectrum rises steeply and at the
long wavelength end can contribute to the extended OH(0,0) band near
3085~\AA. Since the effective area, $A_{eff}$, supplied by the
\GALEX\ project did not extend beyond 3100~\AA, we identified a field
F8 V star and an \IUE\ calibration standard to extend $A_{eff}$ to 3400
\AA.  A model of the dust scattered solar continuum is required to
account for the contamination of the OH(0,0) and CS(0,0) bands.  We
used a solar spectrum from UARS/Solstice, described by
\citet{Woods:1996}, multiplied by this $A_{eff}$, and convolved with a
superposition of $\sim$80~\AA\ and $\sim$250~\AA\ Gaussian instrument
functions to simulate the radially outflowing dust in the dispersion
direction, in order to generate a template for the solar scattered
light.  This template, normalized to regions of the spectrum where
strong molecular emission is absent, and reddened by 4\% per 100~\AA\
\citep{Feldman:1985}, is also shown in the figure.  The
difference between the observed spectra and the template is shown in
the lower panels of the same figure.

The three identified emission features are the CS(0,0) band at
2576~\AA, and the OH(1,0) and (0,0) bands at 2820 and 3085~\AA,
respectively.  For near zero heliocentric velocity, the ratio of
fluorescence efficiencies of the OH(1,0) and (0,0) bands is 0.10
\citep{Schleicher:1988}, but the \GALEX\ effective area is 4.2 times
higher at 2820~\AA\ than at 3085~\AA, so the (1,0) band should give a
count rate $\sim$0.4 that of the (0,0) band.  This is consistent with
the lower panels of Figure~\ref{spec}, and serves to validate the
continuum subtraction.  In principle, we could use the template scaling
factor to derive the temporal variation of the dust ejected by the
impact, but this is much better done with visible observations from the
ground or from space \citep[e.g.,][]{Schleicher:2006, Feldman:2007}
because of uncertainties in the background subtraction.

\placefigure{deep}

\GALEX\ observations began only a few minutes after the impact and
grism spectra were recorded on six successive orbits, or for about 8.5
hours following the impact.  The temporal evolution of the CS and OH
emissions can be seen in the individual spectra shown in
Figure~\ref{deep}.  To construct a light curve, we used a
rectangular aperture wide enough to capture almost all of the band
emission, keeping the aperture height at $19.\!''5$, as above.  For CS,
the width was $22.\!''5$, corresponding to a spectral bandwidth of
100~\AA, while for OH the width was $46.\!''5$, or $\sim$200~\AA.  The
projected sizes on the sky are $12,600 \times 14,500$~km and $12,600
\times 30,000$~km, respectively for CS and OH.  Similar photometry was
performed on the pre- and post-impact data sets, and the results are
shown in Figure~\ref{time}, in which the plotted times are the
mid-points of the exposure.  The observations beginning about 24 hours
after impact show the gaseous emissions returning to near quiescent
levels.

\placefigure{time}

\section{Discussion}

The quiescent production rates were determined using the vectorial model of
\citet{Festou:1981}.  The parent lifetimes at 1~AU were 71,000~s for
\Htwoo, with 85\% of the dissociations producing OH \citep{Budzien:1994},
and 1,000~s for \cstwo\ \citep{Feldman:1999}.  The lifetimes of OH and CS
were taken to be 140,000~s and 100,000~s, respectively, but are not
critical for the aperture size used.  The parent outflow velocity in
the model is $0.85 r^{-1/2}$~\kms, while the daughters are given 1.05
and 0.90~\kms, for OH and CS, respectively.  The fluorescence
efficiencies at 1~AU are $2.3 \times 10^{-4}$ and $5.4 \times 10^{-4}$
photons s$^{-1}$ molecule$^{-1}$ for the OH(0,0) band \citep{Schleicher:1988}
and the CS(0,0) band (R. Yelle, private communication), respectively.  
The derived production rates are given in Table~\ref{prates}, together
with other recently reported values.  Our value for the water
production rate fits squarely within the range reported previously.  We
also note that the \cstwo/\Htwoo\ production rate ratio, $9.5 \times
10^{-4}$, is typical of values found from \IUE\ observations of comets
\citep{Meier:1997}.
 
\placetable{prates}

The linear increase with time following impact in the number of
dissociation products in the aperture can be modeled, using known
photochemical rates \citep{Feldman:2004}, to determine the abundance of the
parent molecular species produced in the collision.  This approach was used
by \citet{Kuppers:2005} for the analysis of wide-field camera images of
OH emission.  However, for our aperture height of 12,600~km, molecules
flowing radially outward at a mean velocity of 0.5~\kms\ will begin to
exit the aperture 3.5~hours after impact.  This problem was considered
by \citet{Manfroid:2007}, and we adopt their formalism.  In addition to
the parent and daughter species, whose photochemical lifetimes are given
above, \citeauthor{Manfroid:2007} also include a ``grandparent''
molecule to account for gas excavation over an extended period (several
minutes) or evaporation from excavated icy grains.  The ``lifetime'' of
the grandparent is an adjustable parameter.

The model results are shown with the data in Figure~\ref{time}.  For a
baseline, we adopt the mean value of pre-impact count rates, which
correspond to an observation almost exactly one rotation period
($\sim$41~hours) earlier.  The CS light curve is well fit by a mean
velocity of 0.4~\kms\ and a velocity dispersion of 0.1~\kms, similar to
the values derived by \citeauthor{Manfroid:2007} for various radicals,
and a ``grandparent'' lifetime of 2000~s.  For OH, models with
0.4~\kms\ and 0.5~\kms\ are shown in the figure.  What matters most in
deriving the number of parent molecules produced is the linear increase
immediately following the impact before any of the dissociation
products has left the field-of-view.  We note that \citet{Feldman:2006}
found a maximum velocity of CO produced by the impact to be 0.7~\kms.

The total number of molecules found is $\sim 1.3 \times 10^{29}$
molecules of \cstwo\ and $\sim 1.6 \times 10^{32}$ molecules of
\Htwoo\ (corresponding to $\sim 1.6 \times 10^4$~kg and $\sim 4.6
\times 10^6$~kg, respectively).  The \cstwo/\Htwoo\ ratio is $8.0
\times 10^{-4}$, comparable to the quiescent value within the given
uncertainties.  A comparison with other reported values is given in
Table~\ref{mols}, and shows generally good agreement.  Our numbers are
lower, by 25\% for \Htwoo\ and a factor of 3 for \cstwo, than those
presented in 2006 in a preliminary report on these results
\citep{Feldman:2009}.  The spectral images used in that analysis did
not have the comet's motion removed and so a large extraction box was
used to ensure that all of the cometary photons were counted.  This
made it difficult to properly subtract the dust continuum, which is more
important for isolating the CS emission than it is for the OH
emission.  We note that the only uncertainties in the derivation of the
total number of molecules produced (other than due to photon statistics
and absolute calibration) are those in the photodissociation lifetimes
and fluorescence efficiencies, which are known at the 10\% level.  The
combined uncertainties are $\approx 20$\%.

\placetable{mols}

After 25 hours all gas and dust emissions within the summation aperture
have returned to nearly quiescent values.  However, as the lower right
panel of Fig.~\ref{spec} shows, the OH bands one day after impact
appear much broader in wavelength than prior to impact, likely due to
outflow along the dispersion direction.  This is not surprising as at
1.5~AU the photodissociation lifetimes of \Htwoo\ and OH are 2 and
3~days, respectively.

Two other orbiting spacecraft made NUV observations of the {\it Deep
Impact} event: both are X-ray observatories with UV/optical imagers
that were used with wide-band NUV filters.  The filter passband of the
{\it Swift} imager included both the CS and OH bands in addition to
dust continuum \citep{Mason:2007} but the photometric images are
dominated by the OH emission.  The number of water molecules produced
by the impact derived from their photometry (see Table~\ref{mols}) is
in excellent agreement with our result.  The {\it XMM-Newton}
observatory imager had two NUV filters, one of which included only CS
emission while the other included both CS and OH \citep{Schulz:2006}.
The light curves derived from both filters show behavior similar to
that of the two light curves in Figure~\ref{time}.
\citeauthor{Schulz:2006} attribute the difference in shape of the light
curves to the presence of small ``icy grains'' in the ejecta.  However,
as the \GALEX\ spectra of Figure~\ref{spec} demonstrate, their data can
be reasonably accounted for by CS emission.

\section{Conclusion}

\GALEX\ observations of comet 9P/Tempel~1 at the time of {\it Deep
Impact} provide a unique measurement of the CS parent produced by the
excavation of cometary material, and as with other daughter molecules,
appears to have the same abundance relative to \Htwoo\ as in quiescent
outgassing of the comet.  Our derived value for the amount of
\Htwoo\ excavated is in excellent agreement with the results of several
other investigations.  This reinforces the idea that nearly all the
volatile species are uniformly mixed with the water ice from the
near-surface layer at which sublimation normally occurs to the tens of
meters to which {\it Deep Impact} excavated material.  This lack of
differentiation from near-surface to tens of meters implies either that
the surface erodes (due to hydrodynamic drag on the dust by the gas) at
about the same rate as the differentiation front proceeds deeper or
that the differentiation has reached depths of many tens of meters
while somehow still leaving the same relative fraction of all the
volatiles throughout the upper tens of meters \citep{A'Hearn:2008}.
Notwithstanding the relative faintness of the comet, the
\GALEX\ results demonstrate the unique capability of wide-field imaging
spectroscopy of comets in the ultraviolet at wavelengths below the
atmospheric cutoff.

\acknowledgments

We thank the \GALEX\ project, and Karl Forster in particular, for their
efforts in planning and executing these observations.  We would like to
acknowledge Tim Conrow and the \GALEX\ pipeline processing team for
helping us to fully utilize the moving target capability of \GALEX,
Jon Giorgini and the JPL HORIZONS team for calculating high precision
comet ephemerides from the perspective of \GALEX, and Wayne Landsman
and the many contributors to the IDLASTRO library.  This work was
supported by the \GALEX\ Guest Investigator program under NASA grant
NNC06GD35G to the Johns Hopkins University.

{\it Facilities:} \facility{GALEX}.



\clearpage

\renewcommand\baselinestretch{1.2}%

\begin{deluxetable}{lccc}
\tablecolumns{4}
\tablewidth{0pc}
\tablecaption{\GALEX\ Grism Observation Log of Comet 9P/Tempel~1. \label{obslog} }
\tablehead{\colhead{FILENAME}&\colhead{EXPTIME (s)}&\colhead{DATE-OBS}&\colhead{TIME-OBS}
}
\startdata
\verb+ GI1_024001_DEEP_IMPACT_PRE_0001-ng-int.fits+    &   820.0 & 2005-07-02 & 06:22:17 \\[-0.025in]
\verb+ GI1_024001_DEEP_IMPACT_PRE_0002-ng-int.fits +   &   940.0 & 2005-07-02  &08:00:53 \\[-0.025in]
\verb+ GI1_024001_DEEP_IMPACT_PRE_0003-ng-int.fits +   &   940.0 & 2005-07-02  &09:39:29 \\[-0.025in]
\verb+ GI1_024001_DEEP_IMPACT_PRE_0004-ng-int.fits +   &   960.0 & 2005-07-02  &11:18:04 \\[-0.025in]
\verb+ GI1_024001_DEEP_IMPACT_PRE_0005-ng-int.fits +   &  940.0 & 2005-07-02  &12:56:41 \\[-0.025in]
\verb+ GI1_024001_DEEP_IMPACT_PRE_0006-ng-int.fits +   &  940.0 & 2005-07-02  &14:35:17 \\[-0.025in]
\verb+ GI1_024001_DEEP_IMPACT_PRE_0007-ng-int.fits +   &  940.0 & 2005-07-02 & 16:13:53 \\[+0.025in] 
\verb+ GI1_024002_DEEP_IMPACT_0001-ng-int.fits +       &  759.0 &
2005-07-04 & 06:01:47  \\[-0.025in]
\verb+ GI1_024002_DEEP_IMPACT_0002-ng-int.fits +       &  959.0 &
2005-07-04  &07:40:23 \\[-0.025in]
\verb+ GI1_024002_DEEP_IMPACT_0003-ng-int.fits +       &  939.0 & 2005-07-04  & 09:18:59 \\[-0.025in]
\verb+ GI1_024002_DEEP_IMPACT_0004-ng-int.fits +       &  938.1 & 2005-07-04  & 10:57:36 \\[-0.025in]
\verb+ GI1_024002_DEEP_IMPACT_0005-ng-int.fits +       &  940.0  &
2005-07-04  & 12:36:11  \\[-0.025in]
\verb+ GI1_024002_DEEP_IMPACT_0006-ng-int.fits +       &  939.0  &
2005-07-04  & 14:14:48  \\[+0.025in]
\verb+ GI1_024003_DEEP_IMPACT_POST_0001-ng-int.fits +  &  299.0 &
 2005-07-05 & 05:05:56 \\[-0.025in]
\verb+ GI1_024003_DEEP_IMPACT_POST_0002-ng-int.fits+   &  919.0 &  2005-07-05  & 06:40:52 \\[-0.025in]
\verb+ GI1_024003_DEEP_IMPACT_POST_0003-ng-int.fits +  &  931.2 &
 2005-07-05 & 08:19:29 \\[-0.025in]
\verb+ GI1_024003_DEEP_IMPACT_POST_0004-ng-int.fits  + & 939.0  &
2005-07-05  &09:58:05  \\
\enddata
\end{deluxetable}

\clearpage

\begin{table}
\begin{center}
\caption{Quiescent production rates (in \mols) in 9P/Tempel~1. \label{prates}}
\medskip
\begin{tabular}{@{}lccc@{}}
\tableline\tableline
Reference & Observation &H$_2$O ($\times$10$^{27}$) & CS$_2$ ($\times$10$^{24}$) \\
\tableline
This paper & OH/CS - UV (\GALEX) & 6.4 & 6.7 \\
\citet{Keller:2007} & OH - UV (\it Rosetta) & 4.2--4.4 & \\
\citet{Schleicher:2006}  &  OH - UV (ground) & 6    & \\
\citet{DiSanti:2007}  & \Htwoo\ - IR & 9.4  & \\
\citet{Bensch:2007} & \Htwoo\ - sub-mm ({\it SWAS}) & 5.2 & \\
\citet{Biver:2007} & \Htwoo/CS - sub-mm ({\it ODIN}) & 9.1 & $< 19.7$\tablenotemark{a} \\
\tableline
\end{tabular}
\tablenotetext{a}{Post-impact on 6 July.}
\end{center}
\end{table}

\begin{table}
\begin{center}
\caption{Total number of water molecules produced by the impact. \label{mols}}
\medskip
\begin{tabular}{@{}lcc@{}}
\tableline\tableline
Reference & Observation &H$_2$O ($\times$10$^{32}$)  \\
\tableline
This paper & OH - UV (\GALEX) & 1.6  \\
\citet{Mason:2007}  &  OH - UV ({\it SWIFT}) & $1.4 \pm 0.2$    \\
\citet{Keller:2007} & OH - UV (\it Rosetta) & 1.5--3.0  \\
\citet{Bensch:2007} & \Htwoo\ - sub-mm ({\it SWAS}) & $<5.9$ (3-$\sigma$) \\
\citet{Biver:2007}  & \Htwoo\ - sub-mm ({\it ODIN}) & $1.7 \pm 0.7$  \\
\tableline
\end{tabular}
\end{center}
\end{table}

\renewcommand\baselinestretch{1.6}%

\clearpage 
\begin{center}{\bf FIGURE CAPTIONS}\end{center}

\figcaption[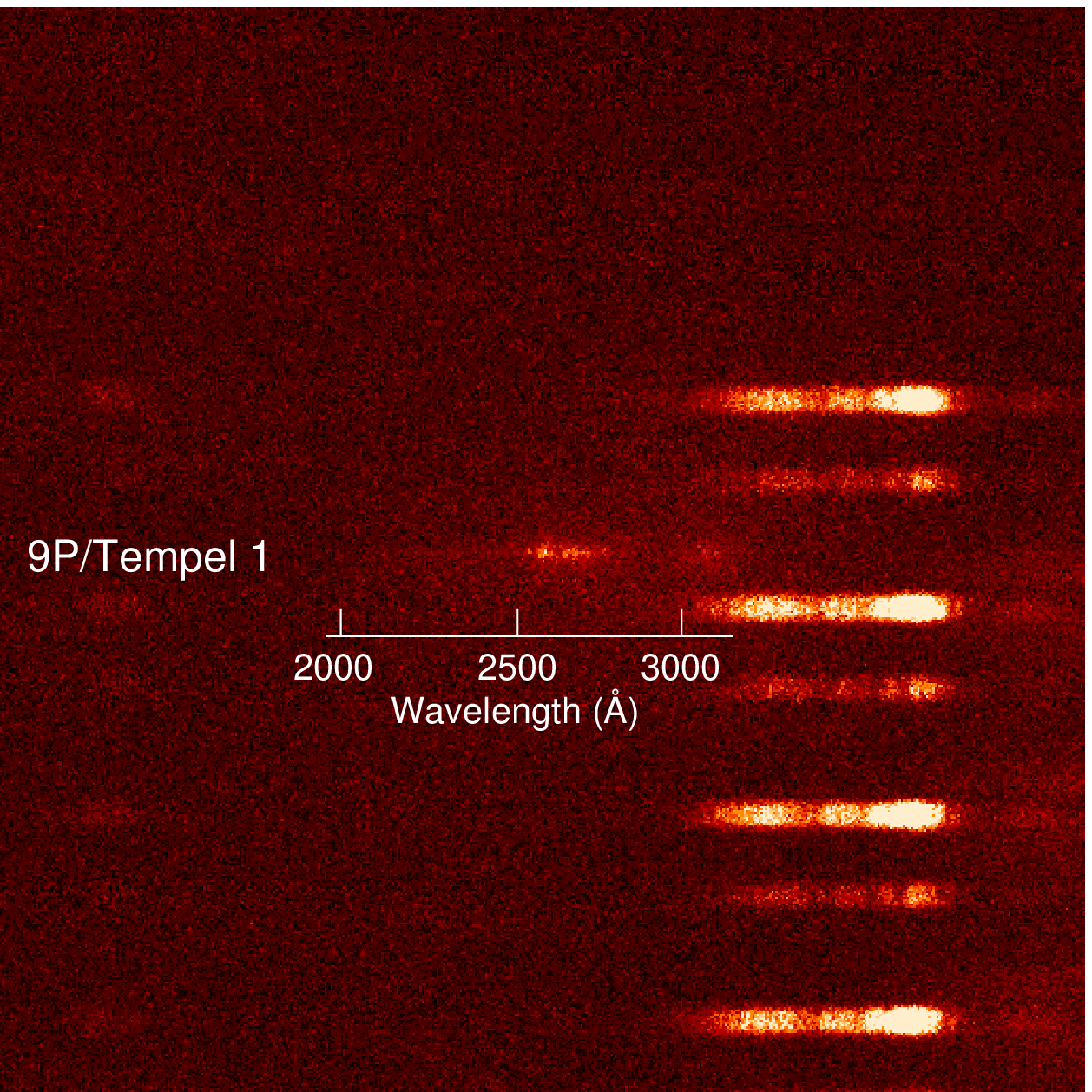]{The central $12.\!'8 \times 12.\!'8$ of the composite 
\GALEX\ NUV grism image taken in the 8.5~h following the {\it Deep Impact}
encounter with comet 9P/Tempel~1.  The motion of the comet is nearly
perpendicular to the dispersion as can be seen from the multiple, wide
stellar spectra.  The image has been rotated 48\degr\ east from north.
Nearly point-like CS and continuum emission together with diffuse OH
emission at 3085~\AA\ can be seen in the spectrum.  \label{image} }

\figcaption[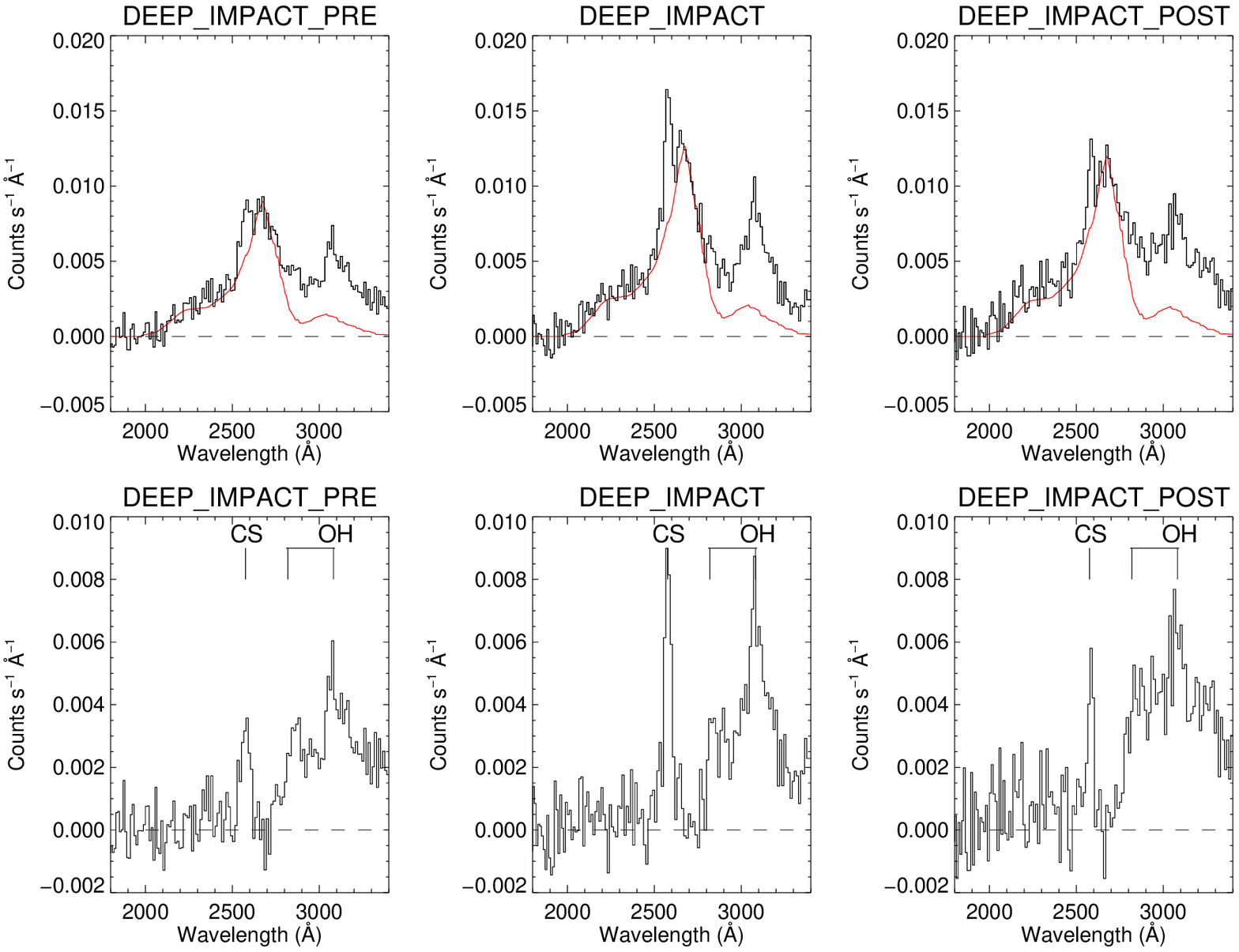]{Composite \GALEX\ NUV spectra of comet
9P/Tempel~1 for each of the three days of observation as listed in
Table~\ref{obslog}.  Background counts have been subtracted.  The top
panels show the fit to solar scattered light in red, while the lower
panels show the spectrum with the solar light subtracted.  The peak
wavelengths of the CS (0,0) and OH (1,0) and (0,0) bands are
indicated.  Note, in the post-impact spectra, the broadening of the OH
emissions due to outflow along the dispersion direction.  \label{spec}}

\figcaption[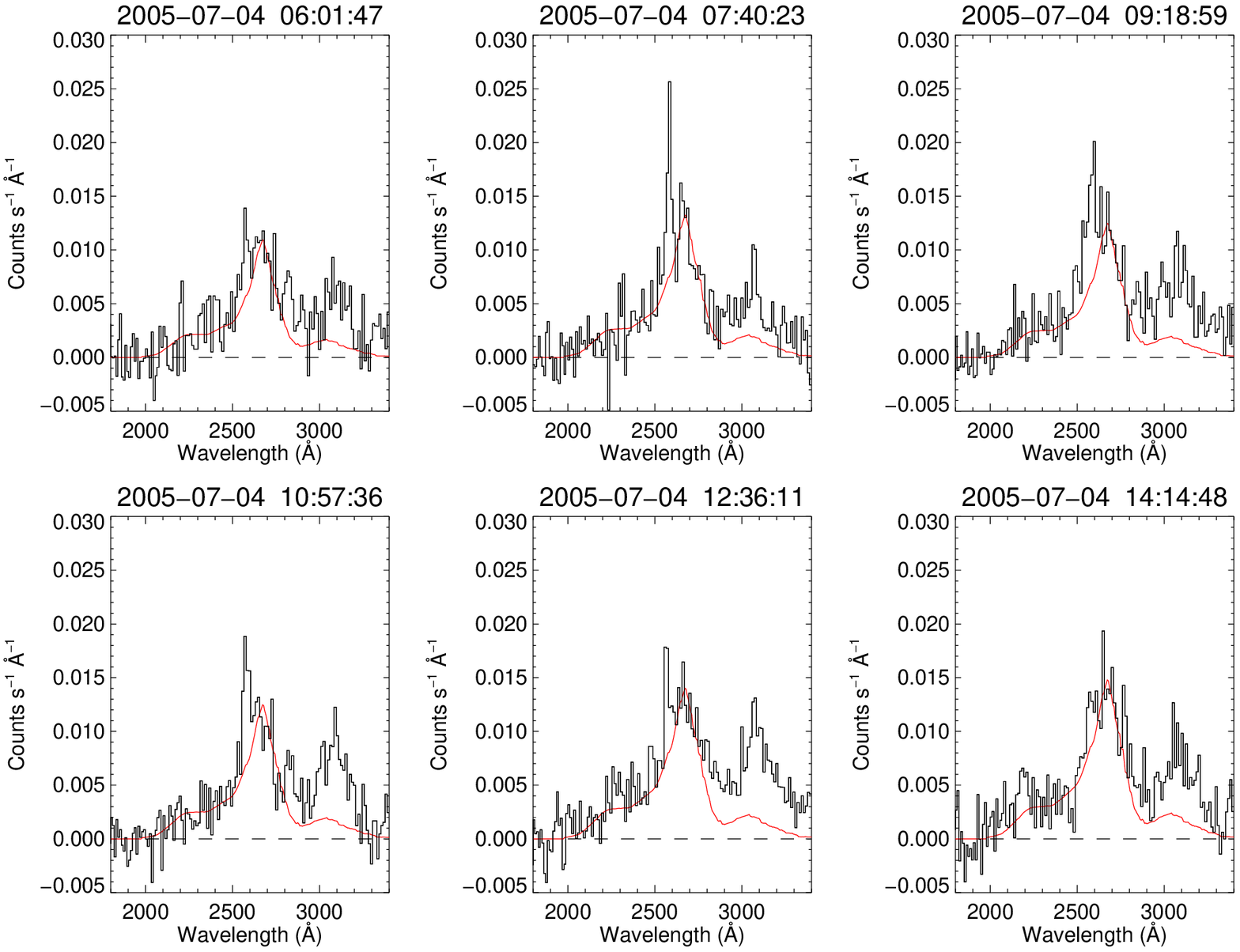]{Same as the top panels in Figure~\ref{spec} for
the six consecutive \GALEX\ orbits taken immediately following the
impact, showing the temporal evolution of the emissions.  \label{deep}}

\figcaption[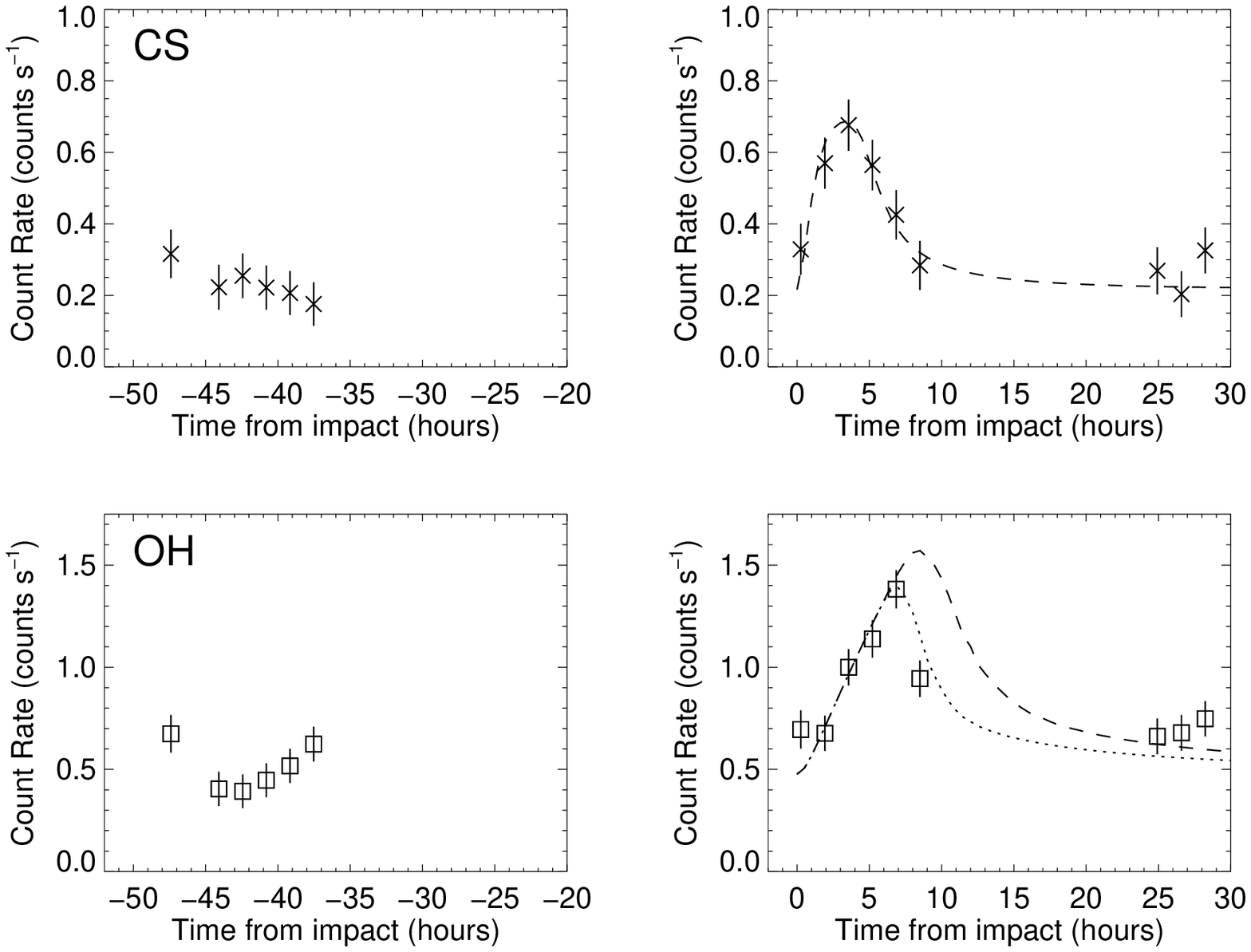]{Temporal evolution of the CS (top) and OH
(bottom) emissions.  The left panels show the count rates roughly one
comet rotation ($\sim$41~hours) prior to the impact.  Model fits to the
aperture photometry are shown for velocities of 0.4~\kms\ (dashed line)
and 0.5~\kms\ (dotted line).  \label{time} }

\clearpage
\setcounter{figure}{0}

\begin{figure*}
\begin{center}
\includegraphics*[scale=1.0,angle=0.]{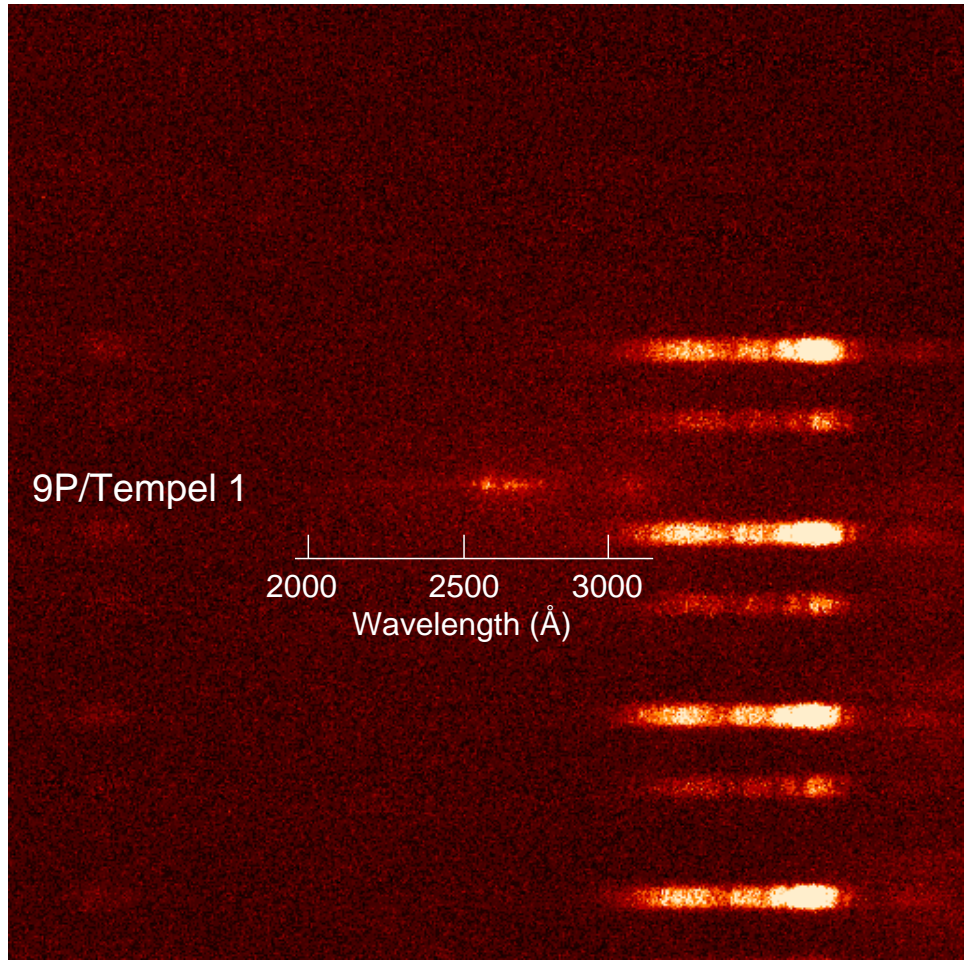} 
\end{center}
\caption[]{The central $12.\!'8 \times 12.\!'8$ of the composite 
\GALEX\ NUV grism image taken in the 8.5~h following the {\it Deep Impact}
encounter with comet 9P/Tempel~1.  The motion of the comet is nearly
perpendicular to the dispersion as can be seen from the multiple, wide
stellar spectra.  The image has been rotated 48\degr\ east from north.
Nearly point-like CS and continuum emission together with diffuse OH
emission at 3085~\AA\ can be seen in the spectrum.  
}
\end{figure*} 

\begin{figure*}
\begin{center}
\includegraphics*[scale=0.9,angle=0.]{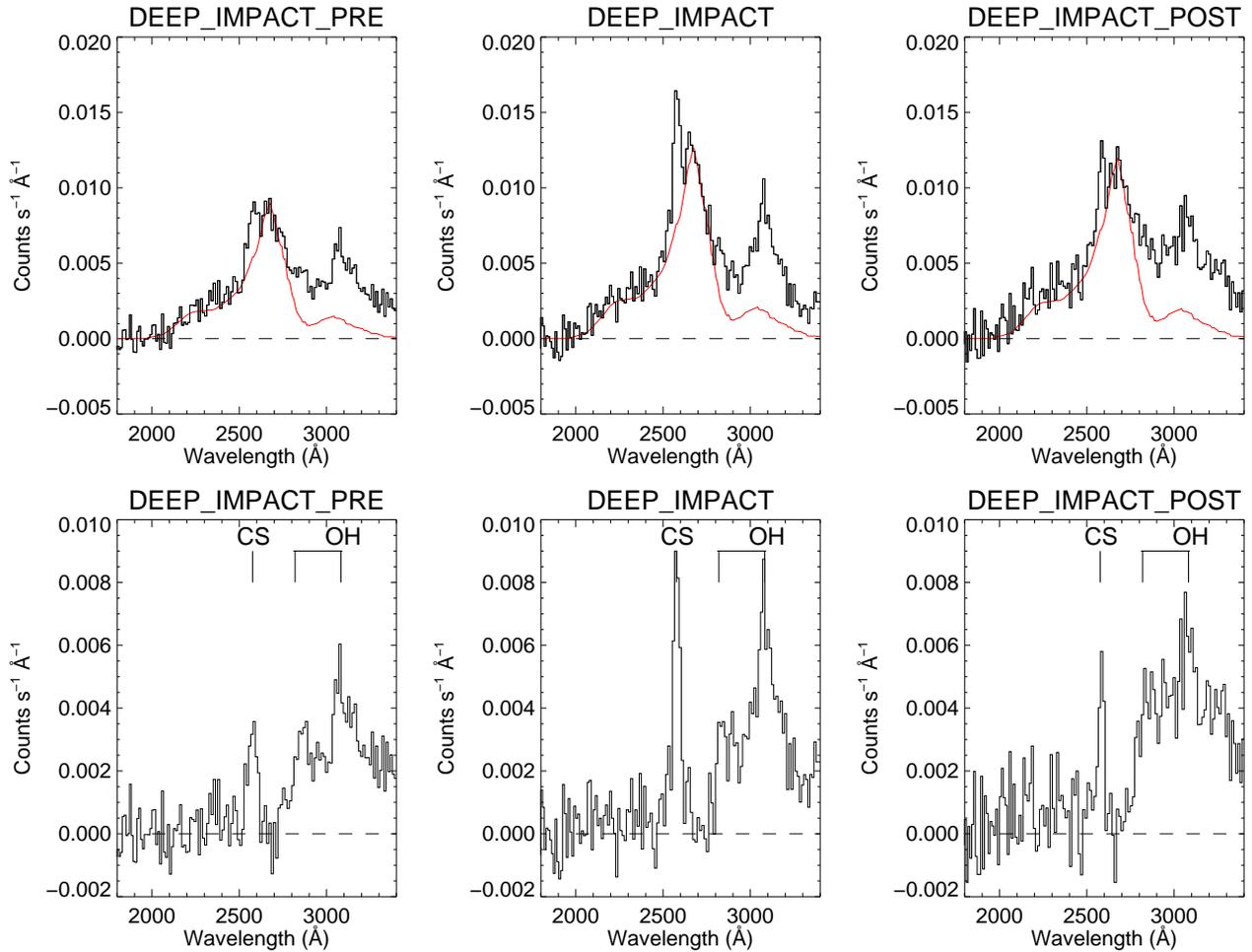} 
\end{center}
\caption[]{Composite \GALEX\ NUV spectra of comet 9P/Tempel~1 for each of
the three days of observation as listed in Table~\ref{obslog}.  Background
counts have been subtracted.  The top panels show the fit to solar scattered
light in red, while the lower panels show the spectrum with the solar
light subtracted.  The peak wavelengths of the CS (0,0) and OH (1,0) and
(0,0) bands are indicated.  Note, in the post-impact spectra, the
broadening of the OH emissions due to outflow along the dispersion
direction.  
}
\end{figure*} 

\begin{figure*}
\begin{center}
\includegraphics*[scale=0.9,angle=0.]{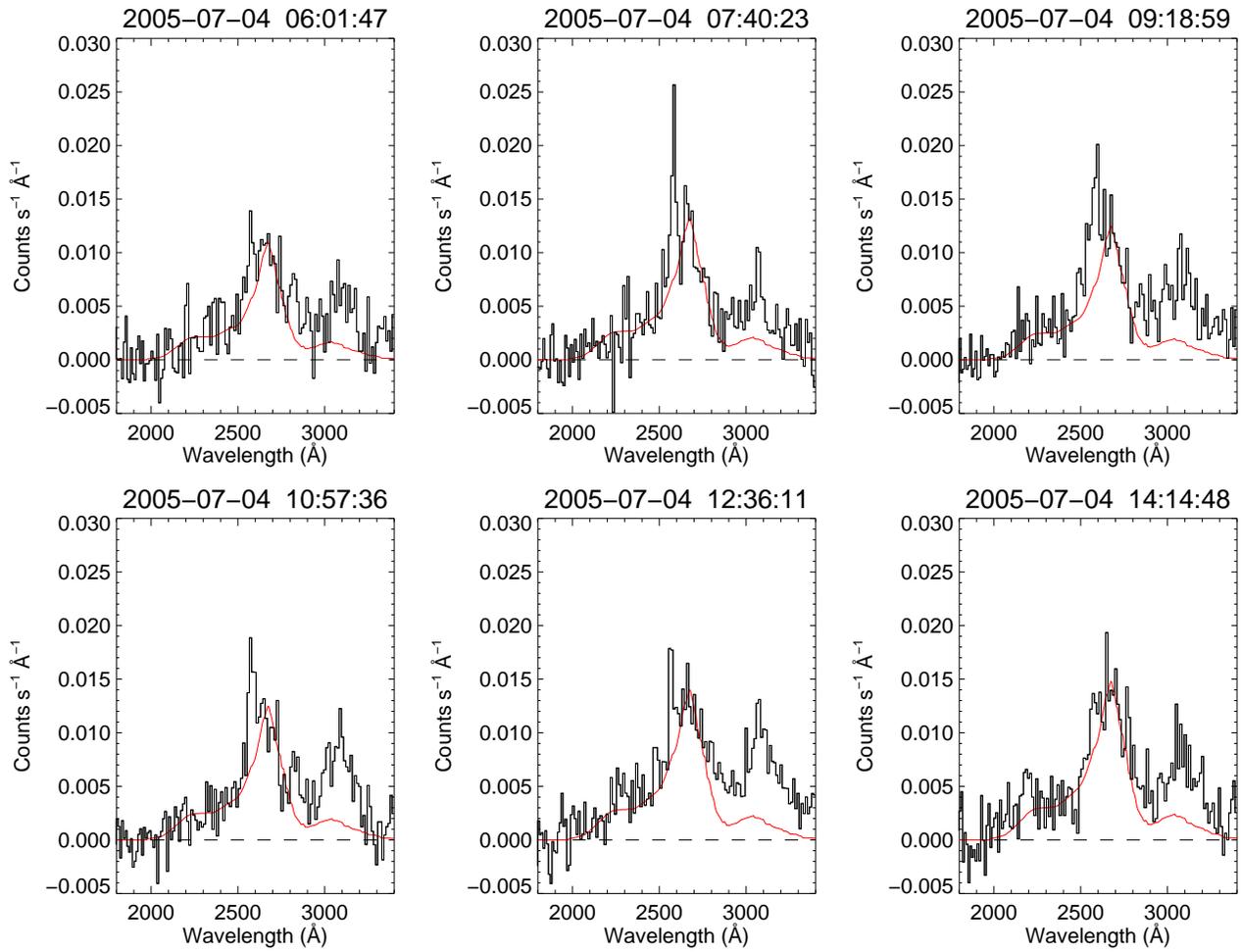} 
\end{center}
\caption[]{Same as the top panels in Figure~\ref{spec} for the six
consecutive \GALEX\ orbits taken immediately following the impact,
showing the temporal evolution of the emissions.  
}
\end{figure*} 

\begin{figure*}
\begin{center}
\includegraphics*[scale=0.90,angle=0.]{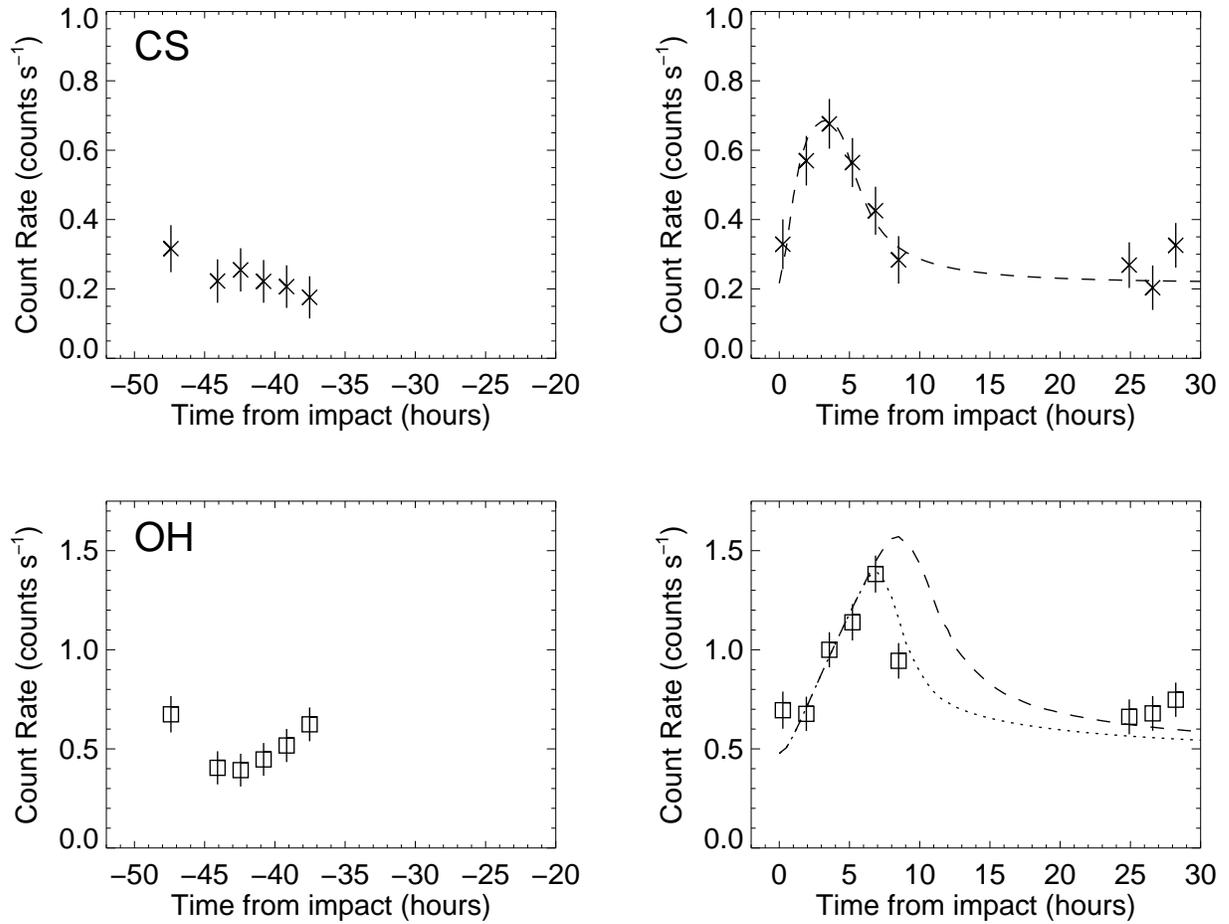}
\end{center}
\caption[]{Temporal evolution of the CS (top) and OH (bottom) emissions.
The left panels show the count rates roughly one comet rotation
($\sim$41~hours) prior to the impact.  Model fits to the aperture
photometry are shown for velocities of 0.4~\kms\ (dashed line)
and 0.5~\kms\ (dotted line).  
}
\end{figure*}

\end{document}